**Update on the multimodal pathophysiological dataset of gradual cerebral ischemia in a cohort of juvenile pigs: auditory, sensory and high-frequency sensory evoked potentials**

Martin G. Frasch[1], Bernd Walter[2,3], Christoph Anders[4] and Reinhard Bauer[3]

[1] University of Washington School of Medicine, Center on Human Development and Disability, Seattle, WA, USA
[2] Department of Spine Surgery and Neurotraumatology, SRH Waldklinikum, Gera, Germany
[3] Institute of Molecular Cell Biology, Jena University Hospital, Jena, Germany
[4] Clinic for Trauma, Hand and Reconstructive Surgery, Division of Motor Research, Pathophysiology and Biomechanics, Jena University Hospital, Jena, Germany

*An extension dataset*

**Corresponding authors:**

Dr. Martin G. Frasch

Department of Obstetrics and Gynecology

University of Washington

1959 NE Pacific St

Box 356460

Seattle, WA 98195

Email: mfrasch@uw.edu

Dr. Reinhard Bauer

Institute of Molecular Cell Biology

Jena University Hospital

Hans Knöll Straße 2

D-07745 Jena, Germany

Email: Reinhard.Bauer@med.uni-jena.de

**Abstract**
We expand from a spontaneous to an evoked potentials (EP) data set of brain electrical activities as electrocorticogram (ECoG) and electrothalamogram (EThG) in juvenile pig under various sedation, ischemia and recovery states. This EP data set includes three stimulation paradigms: auditory (AEP, 40 and 2000 Hz), sensory (SEP, left and right maxillary nerve) and high-frequency oscillations (HFO) SEP. This permits derivation of electroencephalogram (EEG) biomarkers of corticothalamic communication under these conditions. The data set is presented in full band sampled at 2000 Hz. We provide technical validation of the evoked responses for the states of sedation, ischemia and recovery. This extended data set now permits mutual inferences between spontaneous and evoked activities across the recorded modalities. Future studies on the dataset may contribute to the development of new brain monitoring technologies, which will facilitate the prevention of neurological injuries.

## Background & Summary

This extension dataset adds the dimension of *evoked* potentials (EPs) to the recently published[1] dataset containing *spontaneous* ten-channel electrocorticogram (ECoG) and electrothalamogram (EThG) activities, accompanied by the data on cerebral blood flow[2], cardiovascular dynamics and metabolites. Specifically, we present the recordings of the sensorimotor and auditory EPs (SEPs, AEPs) during the same conditions of several sedation states, followed by gradual and controlled cerebral ischemia and 60 minutes of recovery as presented for the spontaneous activity.[1]

The model remains to be that of juvenile pigs where we originally introduced the basic stereotaxic approach to chronically recording EThG and quantifying the effects of isoflurane and fentanyl sedation on the brain electrical activity from a ten-channel ECoG, EThG, and the cerebral blood flow[2], followed by the characterization of the effects of gradual propofol sedation on these parameters.[3]

As for the study of spontaneous brain electrical and cardiovascular activities, for the study of EPs the choice of this animal model is also dictated by its excellent amenability to complex stereotactic chronic instrumentation, including the well-controlled application of SEPs via snout electrodes and AEP via in-ear headphones, prolonged stimulation required for high-frequency oscillations (HFO) SEPs[4,5], studies of sedation and clinically relevant patterns of hypoxic/ischemic injury in a relatively large and gyrencephalic brain.[2]

We hope this dataset will contribute to the exciting area of ongoing research into the physiology of SEPs, HFO SEPs,[6] and mid-latency AEPs[7,8] as manifestations of thalamocortical or intracortical communication that can be harnessed as biomarkers of sedation, cerebral ischemia and neuronal recovery.

For the technical validation, we present the complete raw and processed SEP, HFO SEP and AEP datasets as well as a representative analysis.

We hope the present EP dataset will contribute to development of new brain monitoring technologies, which will facilitate the prevention of neurological disorders.

All experiments were carried out in accordance with the European Communities Council Directive 86/609/EEC for animal care and use. The Animal Research Committee of the Thuringian State government approved laboratory animal protocols.

# Methods

## Instrumentation
**General instrumentation.**

The general surgical procedure is identical to the previously reported approach.[1]

**Instrumentation of the head (Fig. 1).**

The head instrumentation procedure is identical to the previously reported approach (Table 1, Figure 1).[1]

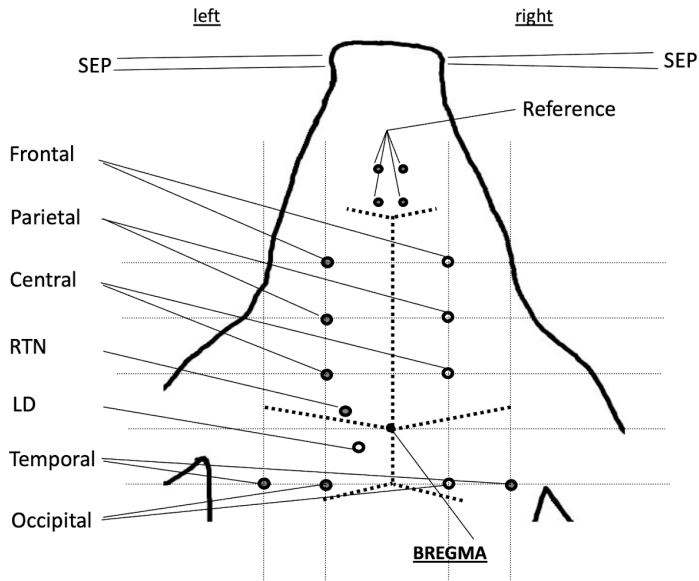

**Figure 1. Instrumentation of the head.**

In addition, the following steps were performed to enable SEP and AEP studies.

SEP electrodes were installed bilaterally on the outer edges of the pig snout to stimulate maxillary nerves.

For AEP studies, the animal's ears were first freed from the stereotactic apparatus. Then, a miniature in-ear headset was inserted on each side (Sennheiser MX 300).

## Description of experimental stages

The experimental protocol is summarized in Figure 2, as presented in [1].

After post-surgical recovery, the pigs were allowed 90 minutes to stabilize (**State 1**) with electrocorticogram (ECoG) and electrothalamogram (EThG) recorded continuously until necropsy. Then, isoflurane in $N_2O$ and $O_2$ was discontinued and ventilation with 100% $O_2$ was performed for 5 minutes. Another phase (**State 2a**) ensued in which intravenous bolus injection of fentanyl (0,015 mg/kg b.w.) was carried out followed by continuous iv infusion of fentanyl (0.015 mg/kg b.w./h) for 90 minutes (**State 2b**). Next, individual doses of propofol required for the maintenance of deep anesthesia were determined under continuous control of mean ABP (MABP). Propofol was infused intravenously (0.9 mg/kg BW/min for ~ 7 min) until a burst suppression pattern (BSP) appeared in the ECoG. The depth of anesthesia was subsequently maintained for 25 minutes via propofol administration (~ 0.35 mg/kg b.w./min) (**State 3**). Next, 30% of the propofol dose required for BSP induction was continuously administered over the course of 90 minutes to produce moderate anesthesia. About ten minutes after the onset of the moderate anesthesia period, the first measurement was performed (**State 4**). **State 5** represented the measurement 60 min after induction of the moderate propofol sedation.

We induced gradual cerebral ischemia as follows.[9] First, the cisterna magna was punctured by a lumbar puncture needle that was fixed in place by dental acrylic resin for elective artificial cerebrospinal fluid infusion/withdrawal to control ICP. Then, the mean ABP was adjusted to about 90 mmHg by the appropriate curbing of the pulmonary trunk diameter with the plastic-coated cerclage. The cerebral perfusion pressure (CPP) was then decreased at 25 mmHg, which was calculated as the difference between MABP and the intracranial pressure (ICP) by appropriate elevation of the ICP. The increase in the ICP was achieved by the infusion of artificial cerebrospinal fluid (warmed to 37°C) into the subarachnoid space via the punctured cisterna magna. The Cushing response during severe brain ischemia was prevented by the appropriate curbing of the pulmonary trunk diameter with the plastic-coated cerclage to control cardiac output. Finally, the cerclage was opened completely, and the artificial cerebrospinal fluid was withdrawn to reach an ICP < 10 mmHg.

The states of gradual ischemia were maintained for 15 min twice (**States 6 and 8**), interceded by a recovery state (**State 7**) lasting 30 minutes in all animals (P746 - P794) except P739 where it lasted 15 minutes. This was followed by 60 min of recovery (**States 9-12**).

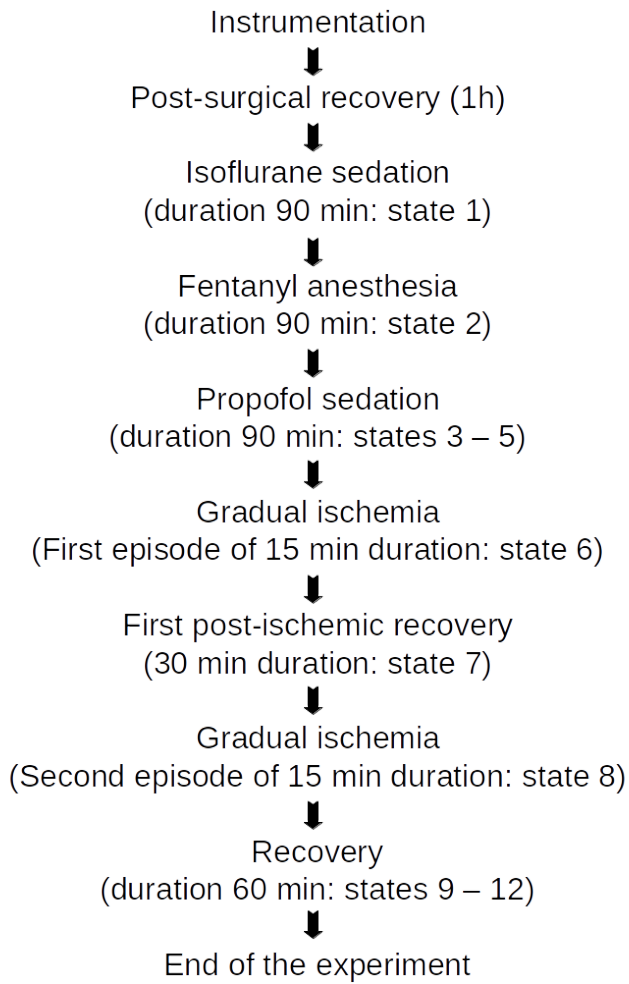

**Figure 2. The experimental protocol.**

On the onset of each state, the evoked potentials were administered in the following order:

1. SEP left
2. SEP right
3. AEP 40 Hz
4. AEP 2000 Hz
5. HFO SEP required up to 16 minutes stimulation duration and, hence, were not performed for each state due to limitations in time allotted for each state.

Table 2 reviews which recordings are available for each state.

**Data acquisition and analyses.**

The approach is identical to what we previously reported.[1]

In addition, the following methods were used to trigger and record SEP and AEP.

As reported in [2], SEPs were induced by bipolar application of constant current rectangular impulses (70 ms, 5 mA applied at 1 pulse/s (HES-Stimulator T, Fa. Hugo Sachs Elektronik, Hugstetten, Germany), for 2 minutes at every side, respectively. The recordings were sampled at 2 kHz and averaged across 100 sweeps.

AEPs were induced by applying auditory stimuli first at 40 Hz and then at 2000 Hz. The auditory stimuli were administered at 90 dB, 150 µs DC impulses, 40 ms duration followed by 5/s frequency with filter bandpass of 15-1500 Hz. The recording electrodes were placed bilaterally on the mastoid; the reference was placed on the vertex. These stimulation parameters were generated using a compiled version of the custom Matlab script provided by Dr. Akeroyd's lab (available for download at the referenced permanent URL).[10,7,11]

HFO SEPs were induced by bipolar application of constant current rectangular impulses (70 ms, 5 mA applied (HES-Stimulator T, Fa. Hugo Sachs Elektronik, Hugstetten, Germany), at a frequency of 5 Hz or 9 Hz for 16 min or 10 minutes at every side, respectively. Raw data files were band-pass filtered between 400 Hz and 800 Hz with an FIR digital filter according to [4] and averaged.

Raw files with constant current rectangular impulse stimulation displayed frequently short-term impulse-synchone artifacts (duration 1 ms at the stimulus time point) with detectable alteration on FIR digital filter effects. Therefore, for each ECoG and EThG channel we removed 4 ms before and after every stimulus and replaced the missing data by a moving average between the previous and subsequent values.

**Signal analysis methodology**

Evoked potentials were calculated using the EP-plugin of Watisa® software. Therefore, recorded stimulus signals were used for event triggering. Subsequently, signal segments of 100 ms duration, starting at the trigger time point, were averaged and stored. Finally, respective EPs determined under identical experimental conditions were averaged as grand mean and used for further analysis.

## Data records

A representative recording is shown in Figure 3.

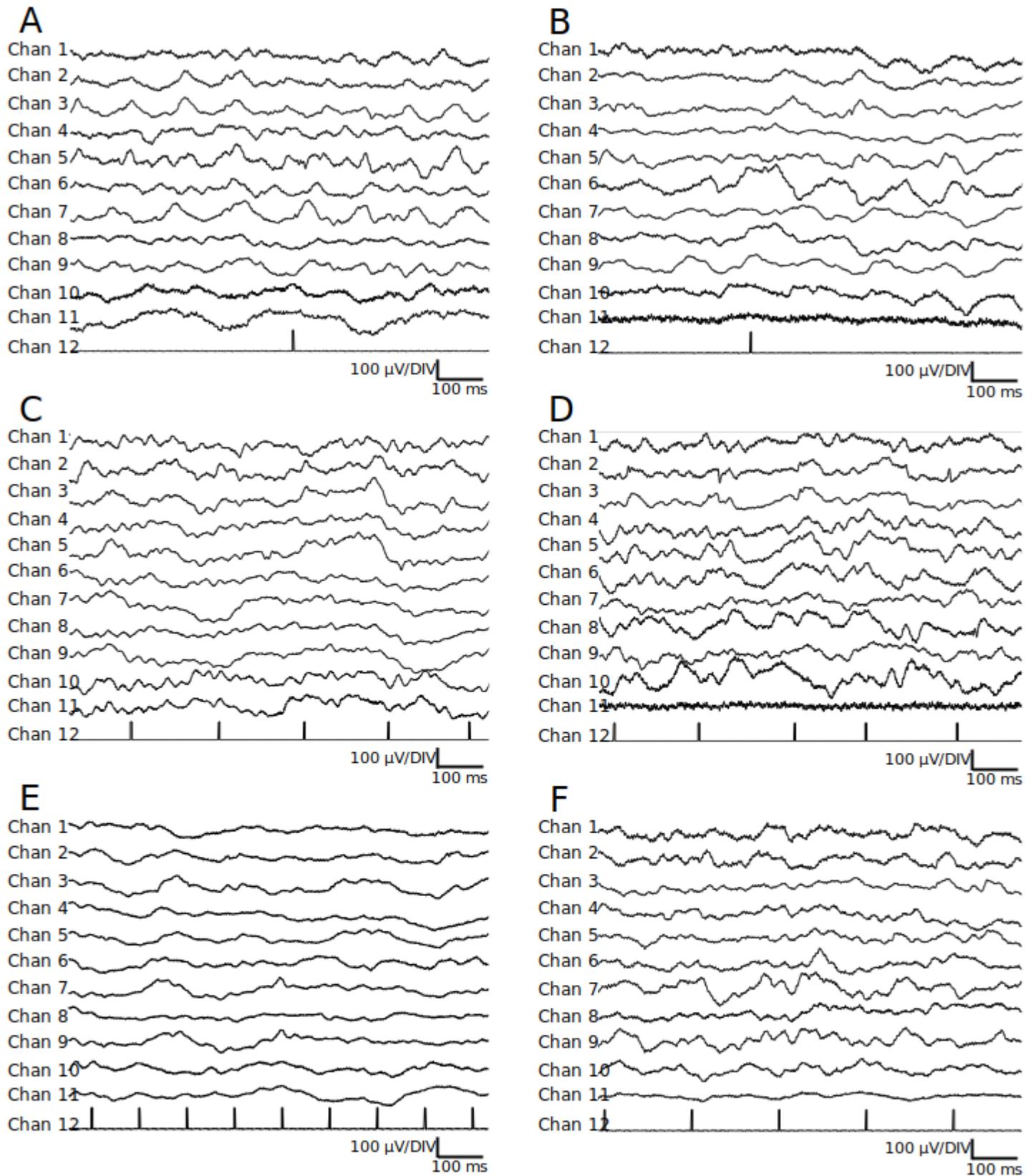

**Figure 3**. Representative raw recording during data acquisition in Watisa® software (State 1).
A: left-side sensorimotor EP stimulation, B: right-side sensorimotor EP stimulation, C: auditory EP stimulation with 40 Hz, D: auditory EP stimulation with 2 kHz, E: sensorimotor EP stimulation with 9 Hz repetition frequency for high frequency oscillation SEP analysis, F: sensorimotor EP stimulation with 5 Hz repetition frequency for high frequency oscillation SEP analysis.

Twelve channels are given at a sampling rate of 2000 Hz (Chan 1, frontal left; Chan 2, frontal right; Chan 3, parietal right; Chan 4, central left; Chan 5, central right; Chan 6, temporal left; Chan 7, temporal right; Chan 8, occipital left; Chan 9, occipital right; Chan 10, RTN nucleus ; Chan 11, LD nucleus; Chan 12, EP stimulus chanel).

The data structure and annotation are as follows. 12 channels are given at a sampling rate of 2000 Hz and a recording duration of about 300 s per state each containing the following channels.

Channels 1-9        ECoG, frontal, parietal, central, temporal, and occipital regions
                       (left parietal ECoG was removed due to artifacts, as before)
Channels 10, 11     EThG from RTN and LD nucleus
Channel 12         SEP or AEP triggers, respectively

Fig. 1 and Table 1 show the electrode arrangement. ECoG channel 3 was of poor quality in most instances due to intermittent technical problems (random faults) and an amplifier breakdown. We hence removed this channel from the dataset presented. That means that the left parietal ECoG (channel 3) is not included in the dataset.

All data has been deposited on FigShare in BIDS format.[12]

## Technical validation

We present the approach and findings of SEPs and AEPs as well HFO SEPs in the ECoG and EThG under the states of sedation, ischemia and recovery.

The SEPs and AEPs are presented as grand means. The complete results are available in the accompanying repository on FigShare.[12] Here we provide the representative findings for sedation, ischemia and recovery states as the grand means (Fig. 4).

For SEPs, we present the first 100 ms which corresponds to the timeframe of the ipsilaterally elicitable early and contralaterally triggered mid-latency SEPs (Fig. 4). Mid-latency SEPs (e.g., N20, P20) allow simultaneous study of the temporally overlapping HFO SEPs.[5,6,13] Consequently, for representative purposes here, we present the Channels 4 and 5 (central ECoG corresponding to C3 and C4) respectively contralateral to the stimulation site. SEP and the corresponding HFO SEP are shown. Note that HFO 5 Hz stimulation only (and not 9 Hz) appears to show correspondence to the mid-latency oscillations of SEP. We observe N20 and P20 components and their HFO counterparts for the first two states of isoflurane and fentanyl sedation, respectively. As noted, not all HFO SEP recordings could be obtained due to the limitations of the required stimulation duration. In the conventional SEPs, the clear modulatory impact of the different sedation modes and the abolition of the response following cerebral ischemia are apparent.

For AEPs, we also focused on the first 100 ms which encompass the brainstem (V), mid-latency (MLAEP, Na, Pa, Nb, Pb) thalamically generated responses, early cortical AEPs (EAEP, from primary auditory cortex) and late cortical responses (LAEP, from frontal cortex and the association areas) (Fig. 5). Overall, our findings agree with literature.[7] Note the mid-latency peak under isoflurane (state 1) which is abolished under fentanyl (state 2), slightly restored under moderate propofol sedation (state 5) and again extinguished 15 and 60 minutes after cerebral ischemia (states 9 and 12 respectively). Temporal ECoG signal shows a more dynamic response overall than central ECoG, as can be expected, with some frequency specificity and partial (40 Hz) or complete recovery (2000 Hz) 60 minutes post ischemia.

The grand means represent the averages of all evoked potential measurements per ECoG/EThG channel per state. The grand means and the underlying individual averaged EP data are available on FigShare and are named and structured as follows:

- "AEP-40Hz"
- "AEP-2000Hz"
- "SEP-left"
- "SEP-right"
- "HFO-SEP"
    - ("HFO-SEP_full_length.xlsx" are HFO SEP grand means where the entire HFO SEP segment was computed and grand means were calculated. The result is then rendered graphically.
    - "HFO-SEP_5-50 ms.xlsx" are HFO SEPs where the above original individual HFO SEP data were cut 0.5-4.5 ms and from 50.5 ms until the end. The result is then rendered graphically. This version removes the initial ~5 seconds 400-800 Hz bandpass FIR swinging artifact on the left side and the higher-order latencies on the right side. The latter may be of interest for future studies on this dataset.

The raw EP data is deposited archived (to retain the file structure) and in the European Data Format (EDF).

"Roadmap for file assessment" file is a spreadsheet that reviews the quality of each recording as we did for the previous dataset.

A total of 11 channels are rendered for the states 1-12 using the same Y-axis settings to ensure the evolution of evoked potentials is captured.
The accompanying file with the table "Number of channels by state and condition" informs how many channels were averaged per state.

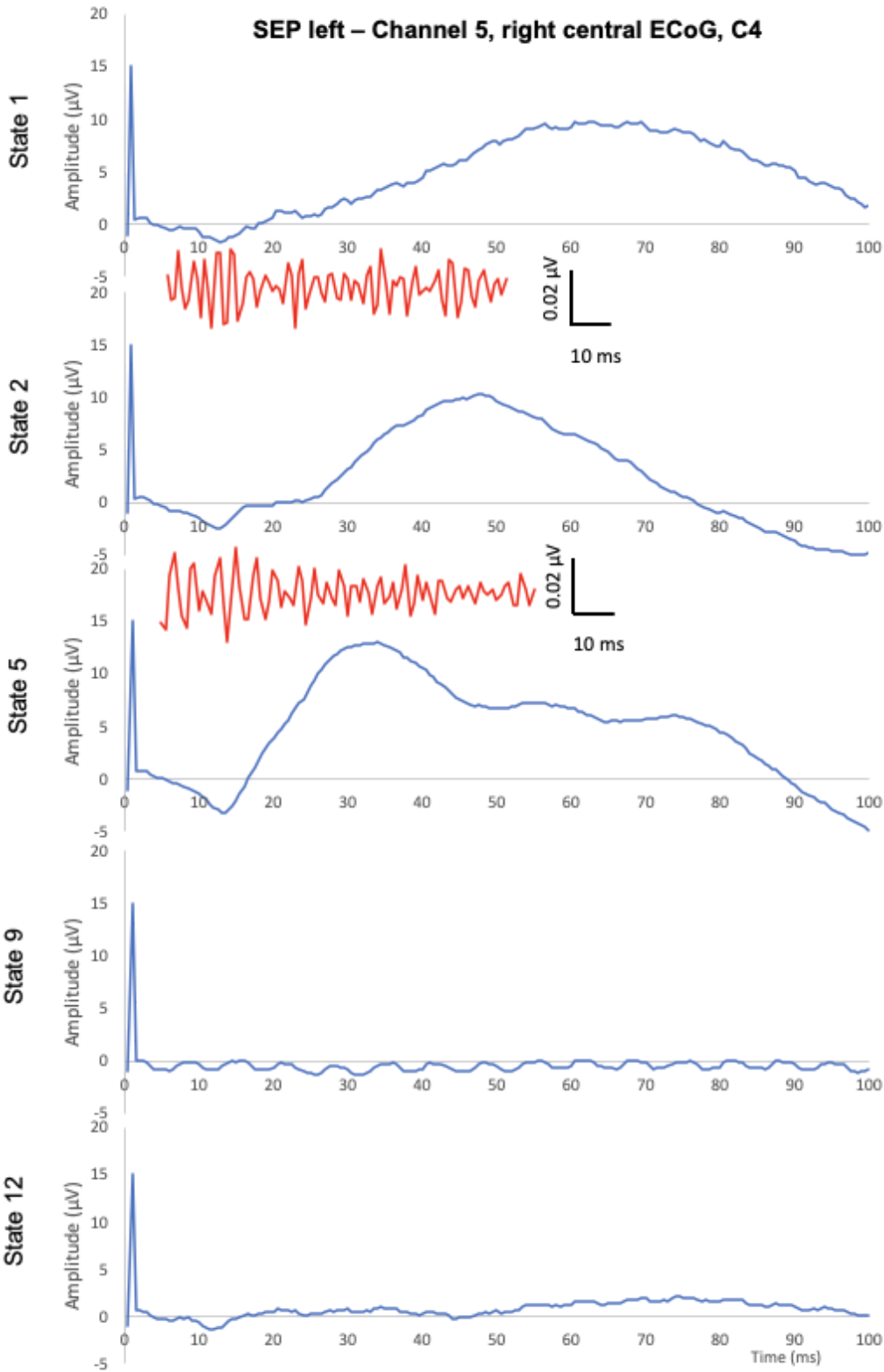

Figure 4A. Left SEPs and HFO SEPs.

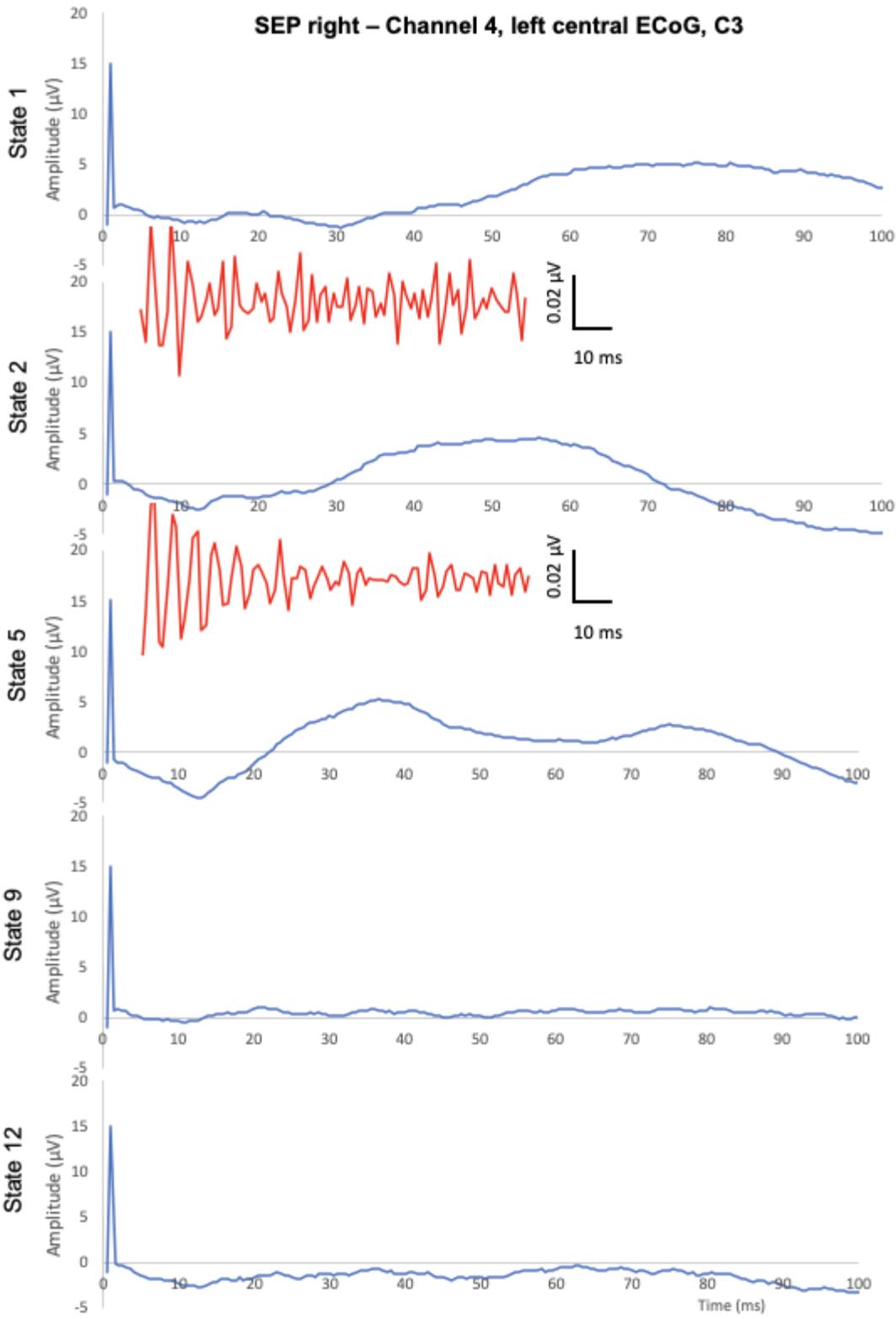

Figure 4B. Right SEPs and HFO SEPs.

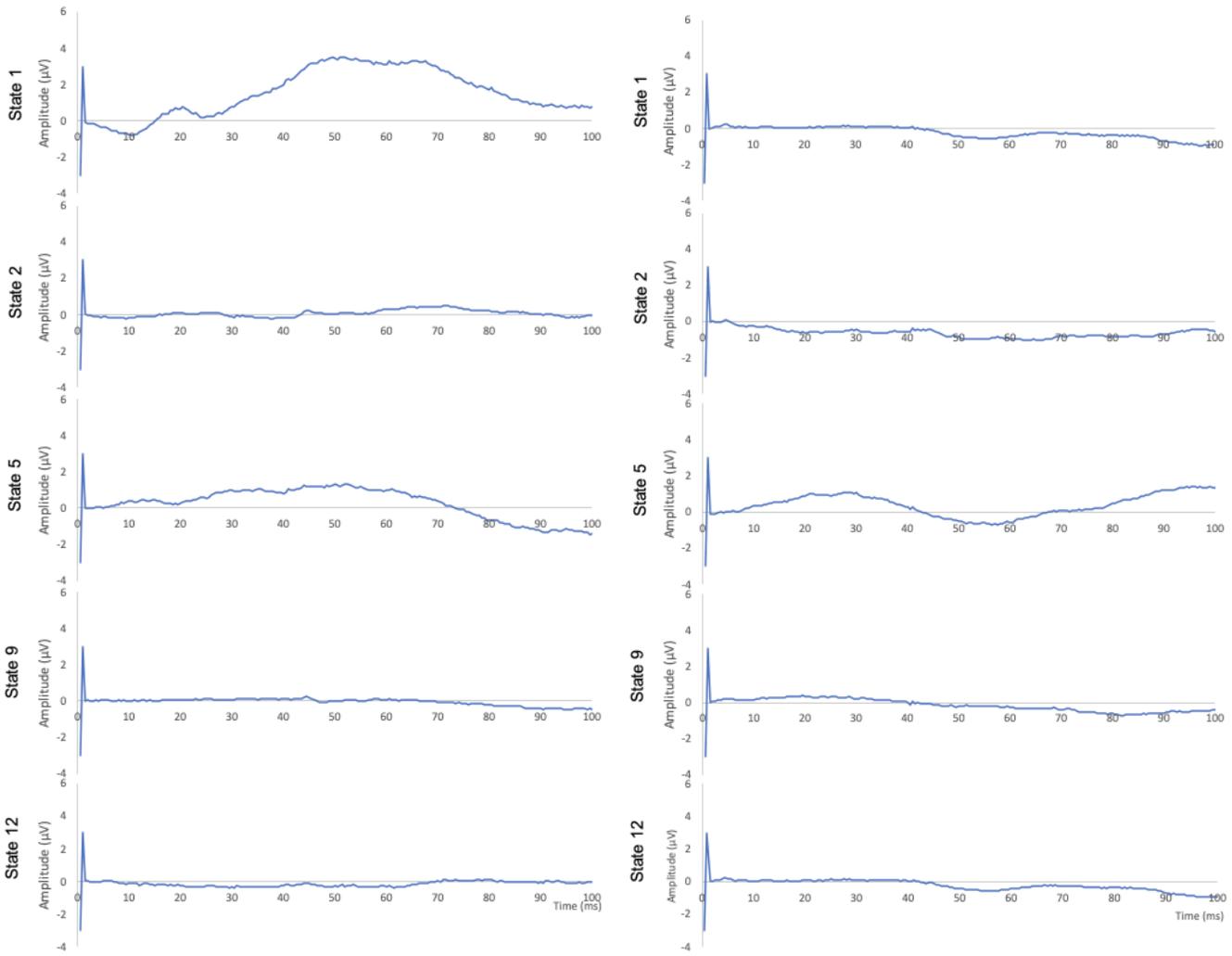

**Figure 5A. AEP.** Central ECoG (right, C4). Left: 40 Hz; Right: 2000 Hz.

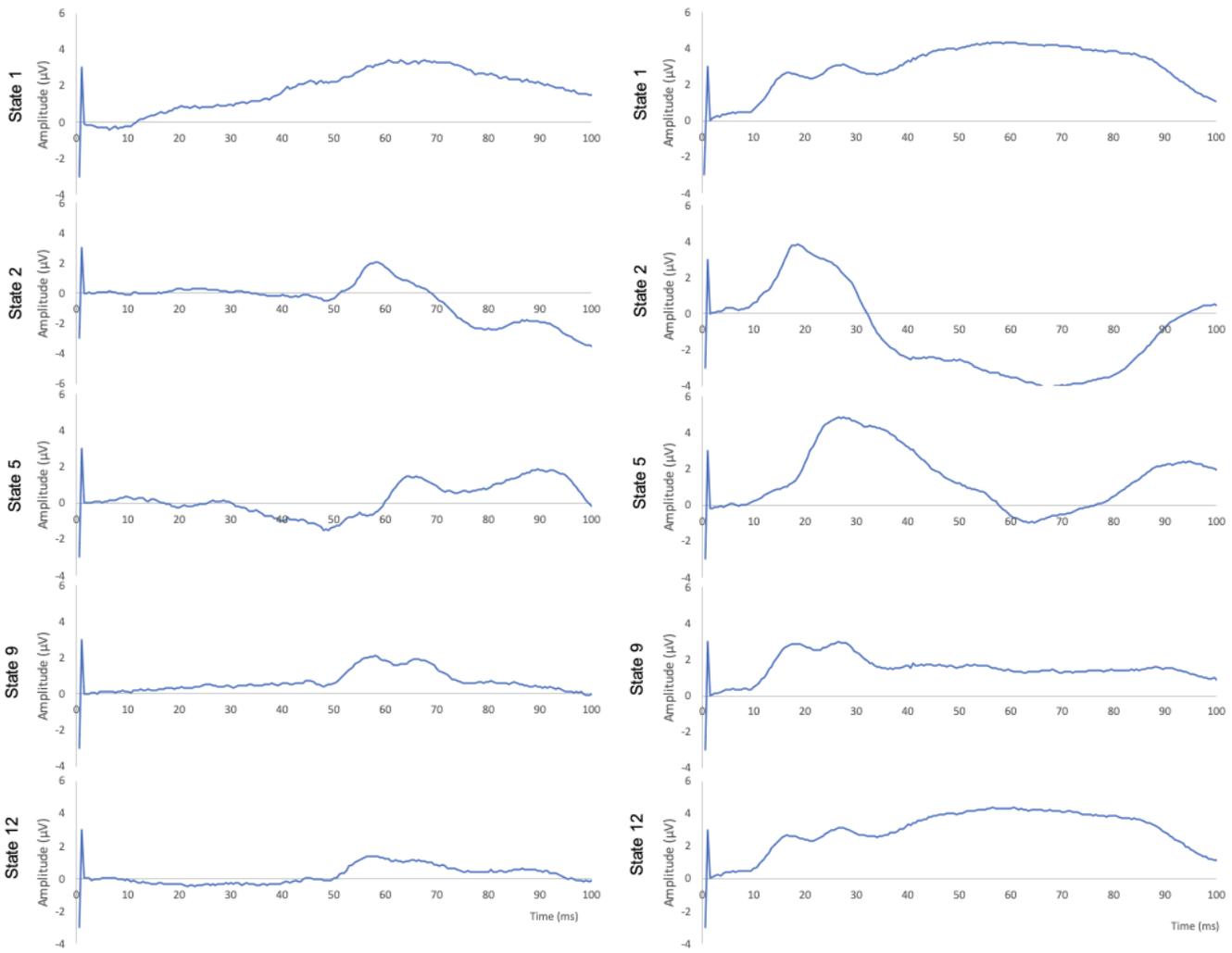

**Figure 5B. AEP.** Temporal ECoG (right, P8). Left: 40 Hz; Right: 2000 Hz.

## Limitations

We recognize that not all EP, in particular HFO SEP, measurements are available for all states and animals. The duration of time intervals between the experimental states left sometimes little room for some of the measurements and we prioritized obtaining the conventional SEP and AEP data over the longer-lasting HFO SEPs. We inventorized the dataset in the roadmap spreadsheet, as part of the FigShare repository, to help navigate the data and design optimal secondary investigations on the data.

## Usage notes

As we reported before[1], the present dataset has been acquired at a 2,000 Hz sampling rate and is hence amenable to studies of the properties of AEPs and SEPs in relation to the HFOs under conditions of various sedation regimes.[14–16] Consequently, thanks to concomitant EThG recordings, it is possible to relate the ECoG patterns of spontaneous HFOs to their thalamic contributions in EThG and their representation in the evoked responses.

## Code availability

EEGLAB has been used which is available as open-source. ERPlab plug-in for EEGLAB can be used as a freely available solution for further EP analysis.

No proprietary code has been deployed in this study except for WATISA (GJB Datentechnik GmbH, Ilmenau, Germany).

## Acknowledgments

The authors thank Konstanze Ernst, Rose Zimmer und Lothar Wunder for their skillful assistance during the experimentation and data analyses. We also gratefully acknowledge the many fruitful conversations with the late Professor Ulrich Zwiener, of good memory, who was an inspiration to us all on this and many other projects.

## Author contributions

R.B. and B.W. conceived, designed, and conducted the experiments and carried out the data analysis. M.G.F. designed and conducted the experiments. R.B., C.A. and M.G.F. carried out the analysis. All authors contributed to the interpretation of the data and drafting of the manuscript. All authors contributed to critical revision and approved the final version of the manuscript.

## Competing interests

The authors have no conflicts of interest to disclose.

## Tables

**Table 1. Stereotactic coordinates* (reproduced with modifications from [1]).**

| Position | Lateral of sagittal suture | Anterior (a) / Posterior (p) of bregma | Depth from the dura | Equivalent according to 10/20 single plane projection of the head[17] |
|---|---|---|---|---|
| Frontal ECoG, Ch. 1 & 2 (left & right) | 12 mm | a 30 mm | | Fp1, Fp2 |
| Parietal ECoG, Ch 3 (left[#] & right) | 12 mm | a 20 mm | | F3[#], F4 |
| Central ECoG, Ch. 4 & 5 (left & right)[$] | 12 mm | a 10 mm | | C3, C4 |
| Temporal ECoG, Ch. 6 & 7 (left & right) | 24 mm | p 10 mm | | T5=P7, T6=P8 |
| Occipital ECoG, Ch. 8 & 9 (left & right) | 12 mm | p 10 mm | | P3, P4 |
| EThG, Ch. 10 (RTN = Nucl. reticularis thalami) | 9 mm | a 2 mm | 24 mm | |
| EThG, Ch. 11 (LD = Nucl. dorsolateralis thalami) | 5 mm | p 2 mm | 20 mm | |

* Reference: Nz
[#], excluded from the dataset due to poor signal quality
[$] channels of particular interest for HFO SEP study[6]

**Table 2. Review of experimental stages and the respective available data sets***

| State # | Experimental stage | EP measurements | Available data sets |
|---|---|---|---|
| state1 | Isoflurane | SEP, AEP | P_728, P_737, P_738, P_739, P_743, P_746, P_749, P_752, P_753, P_791, P_794 |
| | | HFO SEP | P_743, P_746, P_749, P_752, P_753, P_791, P_794 |
| state2 | Fentanyl | SEP, AEP | P_728, P_737, P_738, P_739, P_743, P_746, P_749, P_752, P_753, P_791, P_794 |
| | | HFO SEP | P_739, P_749, P_753, P_791, P_794 |
| state2.5 | 90 min post-fentanyl | SEP, AEP | P_739, P_749, P_753, P_791, |
| state3 | Propofol | SEP, AEP | P_728, P_737, P_738, P_743, P_746, P_752 |
| state4 | Moderate sedation - immediate measurement | SEP, AEP | P_728, P_737, P_738, P_743, P_746, P_752 |
| | | HFO SEP | P_737, P_738, P_743, P_746, P_752 |
| state5 | 60 min post-moderate sedation | SEP, AEP | P_737, P_738, P_743, P_746, P_752 |
| state5.5 | Fentanyl directly pre-ischemia | SEP, AEP | P_794 |
| state6 | 1st ischemic phase: gradual ischemia | SEP, AEP | P_728, P_737, P_738, P_739, P_743, P_746, P_749*, P_752, P_753, P_791, P_794 |
| state 7a | 15 min recovery post-ischemia (first period) | SEP, AEP | P_739, P_746, P_749, P_752, P_753, P_791, P_794 |
| state8 | 2nd ischemic phase: gradual ischemia | SEP, AEP | P_739, P_746, P_749, P_752, P_753, P_791, P_794 |
| state9 | 15 min recovery post-ischemia | SEP, AEP | P_737, P_743 (after single ischemia period); P_739, P_743, P_746, P_749, P_752, P_753, P_791, P_794 (after second ischemia period) |
| state10 | 30 min recovery post-ischemia | SEP, AEP | P_746, P_749, P_752, P_753, P_791, P_794 |
| state11 | 45 min recovery post-ischemia | SEP, AEP | P_737, P_739, P_743, P_749, P_752, P_753, P_791, P_794 |
| state12 | 60 min recovery post-ischemia | SEP, AEP | P_737, P_743, P_746, P_749, P_752, P_753, P_791, P_794 |

*, Details including any artifacts observed, for each file and for each channel, are provided in the Supplemental spreadsheet, also available on FigShare.[12,18]

**Figure legends**

**Figure 1. Instrumentation of pig for recording ECoG and EThG.**

**Figure 2. Experimental protocol of pig model of sedation and gradual ischemia.**

**Figure 3**. Representative raw recording during data acquisition in Watisa® software (State 1).
A: left-side sensorimotor EP stimulation, B: right-side sensorimotor EP stimulation, C: auditory EP stimulation with 40 Hz, D: auditory EP stimulation with 2 kHz, E: sensorimotor EP stimulation with 9 Hz repetition frequency for high frequency oscillation SEP analysis, F: sensorimotor EP stimulation with 5 Hz repetition frequency for high frequency oscillation SEP analysis.
Twelve channels are given at a sampling rate of 2000 Hz (Chan 1, frontal left; Chan 2, frontal right; Chan 3, parietal right; Chan 4, central left; Chan 5, central right; Chan 6, temporal left; Chan 7, temporal right; Chan 8, occipital left; Chan 9, occipital right; Chan 10, RTN nucleus ; Chan 11, LD nucleus; Chan 12, EP stimulus chanel).

**Figure 4. Representative findings of SEPs and HFO SEPs.**
Left and right SEPs are shown in the respective contralateral ECoG channels 5 and 4 along with the corresponding HFO responses (in red).

**Figure 5. Representative findings of AEPs.**
   A. Central ECoG (right, C4). Left: 40 Hz; Right: 2000 Hz.
   B. Temporal ECoG (right, P8). Left: 40 Hz; Right: 2000 Hz.